# The mechanism of Li deposition on the Cu substrates in the anode-free Li metal batteries

*Genming Lai[1], Junyu Jiao[1,\*], Chi Fang[1], Liyuan Sheng[2], Yao Jiang[3], Chuying Ouyang[3], and Jiaxin Zheng[1,3,\*]*

[1]School of Advanced Materials, Peking University, Shenzhen Graduate School, Shenzhen 518055, People's Republic of China

[2]PKU-HKUST ShenZhen-HongKong Institution, Shenzhen 518055, People's Republic of China

[3]Fujian science & technology innovation laboratory for energy devices of China (21C-LAB), Ningde 352100, People's Republic of China

**Corresponding Author**

\* Junyu Jiao, E-mail: jyjiao@pku.edu.cn

\* Jiaxin Zheng, E-mail: zhengjx@pkusz.edu.cn

**Abstract:** Due to the rapid growth in the demand for high-energy-density Li batteries and insufficient global Li reserves, the anode-free Li metal batteries are receiving increasing attention. Various strategies, such as surface modification and structural design of Cu current collectors, have been proposed to stabilize the anode-free Li metal batteries. Unfortunately, the mechanism of Li deposition on the Cu surfaces with the different Miller indices is poorly understood, especially on the atomic scale. Here, a large-scale molecular dynamics simulation of Li deposition on the Cu substrates was performed in the anode-free Li metal batteries. The results show that the Li layers on the Cu (100), Cu (110), and Cu (111) surfaces are closer to the structures of Li (110), Li (100), and Li (110) surfaces, respectively. The mechanism was studied through the surface similarity analysis, potential energy surfaces, and lattice features. Finally, a proposal to reduce the fraction of the (110) facet in commercial Cu foils was made to improve the reversibility and stability of Li plating/stripping in the anode-free Li metal batteries.

## 1. Introduction

Li metal is regarded as an ideal anode for the next-generation secondary batteries, which affords the lowest electrochemical potential (3.04 V vs. the standard hydrogen electrode) and the highest theoretical specific capacity (3860 mA h g$^{-1}$).[1-6] However, the low coulombic efficiency and potential hazards have restricted their applications.[7-8] To alleviate these problems at a low cost, anode-free Li metal batteries have received increasing attention.[3, 9-11] The architecture of anode-free Li metal batteries is constructed from a fully lithiated cathode with a bare anode Cu current collector (CC).[11-12] In such an anode-free Li metal battery, the energy density can be extended to the maximum limit involving no excess Li.[11] Moreover, the anode-free Li metal batteries have higher cost-effective and safety performance than typical Li metal batteries.[9-10] However, gaining more benefits from the anode-free Li metal battery architecture demands the new challenge of Li plating/stripping reversibility and stability to be met.[9]

The Cu CC acts as a vital role in affecting the Li plating/stripping behavior and determines the cycling performance of anode-free Li metal batteries finally.[13] Many modifications to the Cu CC have been proposed for improving the stability of Li metal batteries in the experiments,[14-16] such as free-standing Cu nanowire network,[14] 3D porous Cu foil,[17] and 3D Cu mesh,[18] etc. In addition, several numerical simulation methods have been introduced to study the CCs and Li deposition. For instance, Vikram *et al.* did a computational screening of potential CCs for enabling anode-free Li metal batteries.[19] Ingeborg *et al.* used density functional theory (DFT) and molecular dynamics (MD) simulations to study how the crystal structure of the Cu CC influences the morphology of the Li anode and the mobility of Li on the anode.[20] In a previous study, we achieved a large-scale MD simulation of Li deposition and found that facilitating Li self-healing and improving the fluidity of the deposited Li atoms could effectively facilitate the formation of dendrite-free Li.[21] However, the Cu CC has not reached its full potential so far.[22] On the other hand, the Miller index of the plane in a cubic metal crystal structure is a crucial factor in affecting the Li plating.[23] For instance, Gu *et al.* demonstrated that the Cu (100) surface is lithiophilic and enables a more uniform Li deposition and higher cycling stability.[24] Ju *et al.* achieved the dendrite-free Li deposition by constructing a novel Cu substrate with sharp wrinkles and a highly uniform (100) facet.[25] These studies highlighted the importance of the morphology and structure of the Cu CC surfaces, and demonstrated that controlling the exposed facets of the Cu CC can effectively improve the Li deposition behavior. Nevertheless, how the Miller indices of the Cu CC surfaces impact the Li deposition behavior and the mechanism of Li deposition on the Cu CC surfaces with the different Miller indices on an atomic level are still remain unknown. Thus, an atomic-scale study on the mechanism of Li deposition on the Cu surfaces with the different Miller indices is required.

Herein, a large-scale MD simulation was carried out by using a deep neural network interface potential for Li-Cu systems (LiCu-NNIP) with quantum-mechanical computational accuracy. Through the above simulation, we studied the dynamic behavior of Li deposition on the Cu surfaces with the different Miller indices and the arrangement features of Li atoms. In addition, we used the surface similarity analysis (SSA) method to analyze the structure of the Li layers on the Cu surfaces, which provides quantitative analysis for the study of surface structure. The simulation results demonstrate that Li deposition on a certain Cu surface exhibits a unique panel. Thus, the surface properties of the Li panel can be altered through the different Cu substrates.

## 2. Results

### 2.1 The features of Li deposition on the Cu substrates

Three surface structures with the different Miller indices, Cu (100), Cu (110), and Cu (111), were constructed for the MD simulations, as shown in the top panels of Fig. 1a–c, respectively. These surfaces are the most commonly exposed crystal planes of Cu materials.[26-27] As shown in the top panels of Fig. 1a,c, during deposition, the Li-Cu interface is extremely flat and the Li layers close to the interface are still in order on the Cu (100) and Cu (111) surfaces. The result suggests that the deposited Li atoms only diffuse on the surfaces and the alloying does not occur on the Cu (100) and Cu (111) surfaces. However, on the Cu (110) surface, some Cu atoms exchange with Li atoms at the interface, as shown in the top panel of Fig. 1b and Fig. 2b,e. The result demonstrates that the alloying exists on the Cu (110) surface during deposition and a disordered Li-Cu interface forms on the Cu (110) surface finally (Fig. S1, Supporting Information).

To investigate the features of Li deposition on the Cu surfaces quantitatively, we used the SSA method to compare the similarity between the structures of Li monolayers and the structures of Li standard crystal surfaces (See the Method section for details). Herein, the undermost ten Li monolayers close to the interface were sliced for performing the SSA method, as shown in the bottom panels of Fig. 1a-c. The results show that the best matching surface structures of Li layers on the Cu (100), Cu (110) and Cu (111) surfaces are the Li (110), Li (100) and Li (110), respectively. Besides, the similarity has a rapid increase from the first layer to the second layer, and the subsequent layer gradually increases or remains unchanged. These results demonstrate that the structures of the first two deposited Li layers are decisive in the subsequently deposited layers.

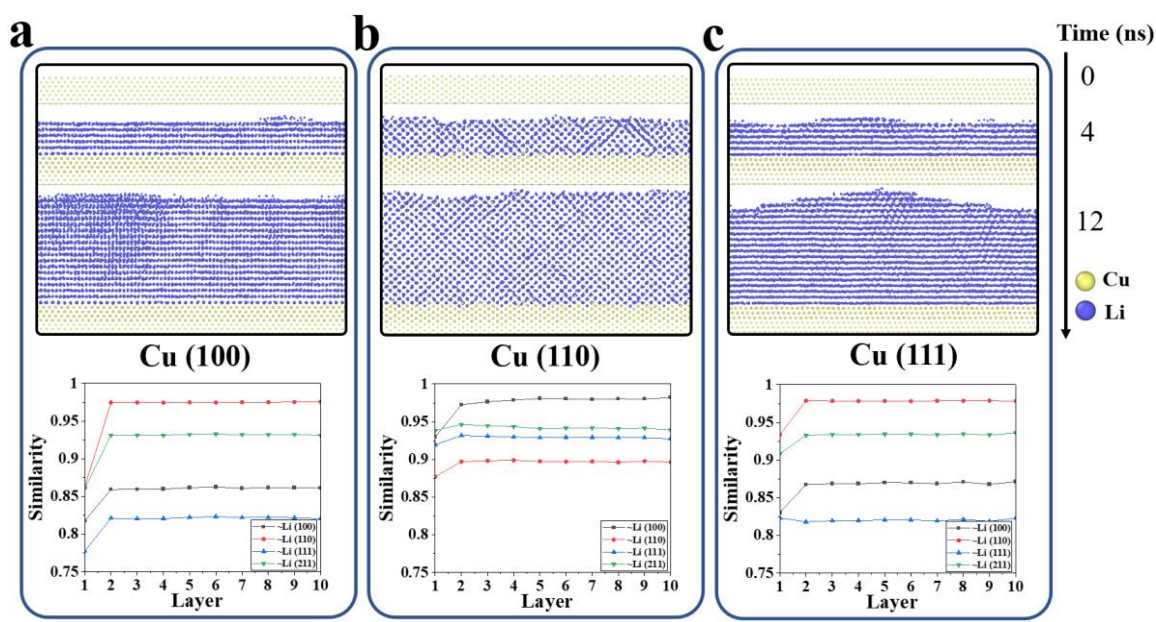

**Fig. 1** Li homogeneous deposition on the Cu surfaces with the different indices and the results of SSA. (a-c) Snapshots of Li homogeneous deposition and the curve of SSA results of the first 10 Li monolayers on the Cu (100) (a), Cu (110) (b), and Cu (111) (c) surfaces.

**2.2 The mechanism of Li deposition on the Cu substrates**

We analyzed the arrangement features of Li atoms on the Cu (100), Cu (110), and Cu (111) surfaces to study the detailed mechanism of Li deposition on the Cu surfaces. Fig. 2a-f are the enlarged view of the center areas (60 Å < x < 70 Å, 15 Å < y < 25 Å) of the Cu (100), Cu (110), and Cu (111) surfaces, respectively. (The diagrams of complete surface structures are shown in Fig. S2, Supporting Information.) They show the atomic arrangements of the first two deposited Li layers on the mentioned surfaces. From Fig. 2a, the first layer Li atoms on the Cu (100) surface are mainly arranged in the center of the quadrilateral structures formed by the underlying Cu atoms. The result indicates that the arrangement of the first layer Li atoms follows the structural features of the Cu (100) surface. The arrangement features of the first layer Li atoms on the Cu (110) and Cu (111) surfaces have a similar pattern, as shown in Fig. 2b,c.

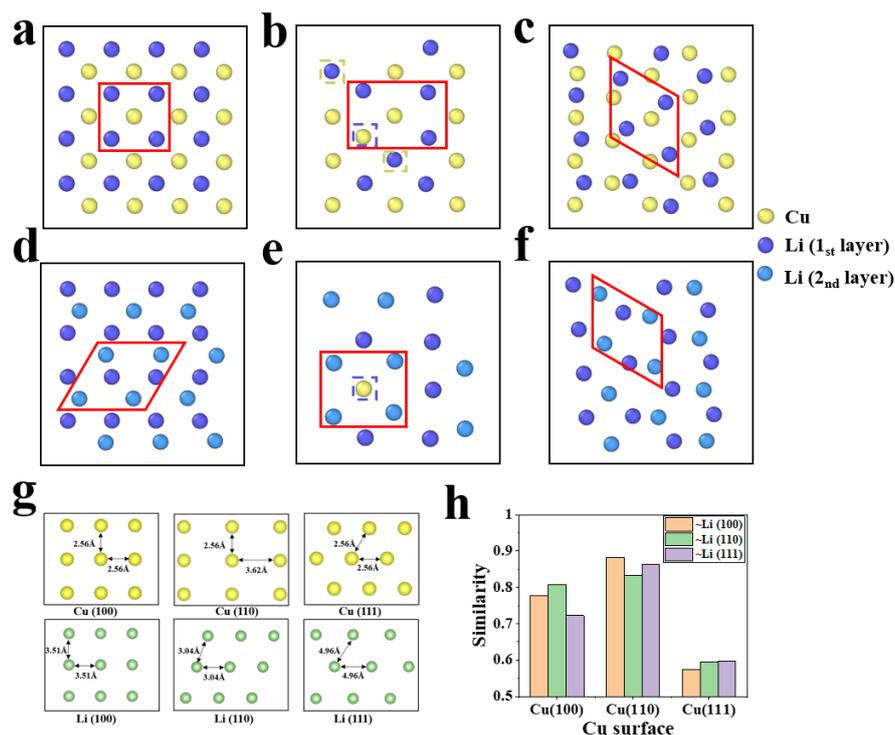

**Fig. 2** Enlarged view of arrangement of partial Li atoms and the atomic arrangement of the standard structure of Cu and Li surfaces. (a-c) The arrangement of the first layer Li atoms at the center of the Cu (100) (a), Cu (110) (b), and Cu (111) (c) surfaces. (d-f) The arrangement of the second layer Li atoms at the center of the Cu (100) (d), Cu (110) (e) and Cu (111) (f) surfaces. (g) The atomic arrangement of the standard FCC Cu (100), Cu (110), Cu (111), BCC Li (100), Li (110), and Li (111) surfaces. (h) The SSA results for the standard structure of the Cu (100), Cu (110), and Cu (111) surfaces.

The potential energy distributions of one Li atom placed on the Cu (100), Cu (110), and Cu (111) surfaces were compared. As shown in the bottom panels of Fig. 3d,e,f, all the Cu surfaces exhibit the local minimum potential energy sites surrounded by the local maximum potential energy sites, periodically. Besides, these local minimum potential energy sites correspond to or close to the center of the quadrilateral structures formed by the underlying Cu atoms (Fig. 3a,b,c). Thus, the arrangements of the first layer Li atoms on these surfaces follow the structural features of the corresponding substrates surfaces.

Starting from the second layer, the deposited Li atoms are no longer arranged along the structural features of Cu substrates, but along the structural features of Li itself, as shown in Fig. 2d–f. Li has a body centered cubic (BCC) structure, and Li (110) is the most stable crystal plane of Li metal.[23-24] It means that the Li atoms are more likely to spontaneously form the structure of Li (110) when Li atoms can move freely without hindrance (Fig. S6-7, Supporting Information). After the Cu surfaces are covered by one layer of Li atoms, the potential energy surfaces on the Li covered Cu (100) and Cu (111) surfaces become flatter, as shown in the top panels of Fig. 3d, f. We calculated the potential energy variances of Li on the different layers, the potential energy variances of Li on the Cu (100) and Cu (111) surfaces decrease after the Cu surfaces are covered by one layer of Li atoms. Thus, the Li atoms on the Li covered Cu (100) and Cu (111) surfaces are no longer trapped in the structure of substrates surfaces. So, Li atoms have the ability to spontaneously form the most stable plane, namely Li (110). The Cu (100) and Cu (111) surfaces have a high lattice match and high similarity with the Li (110) surface, as shown in Fig. 2g,h. Therefore, the second layer Li atoms on the Cu (100) and Cu (111) surfaces are approximately arranged with the structural features of the Li (110) surface.

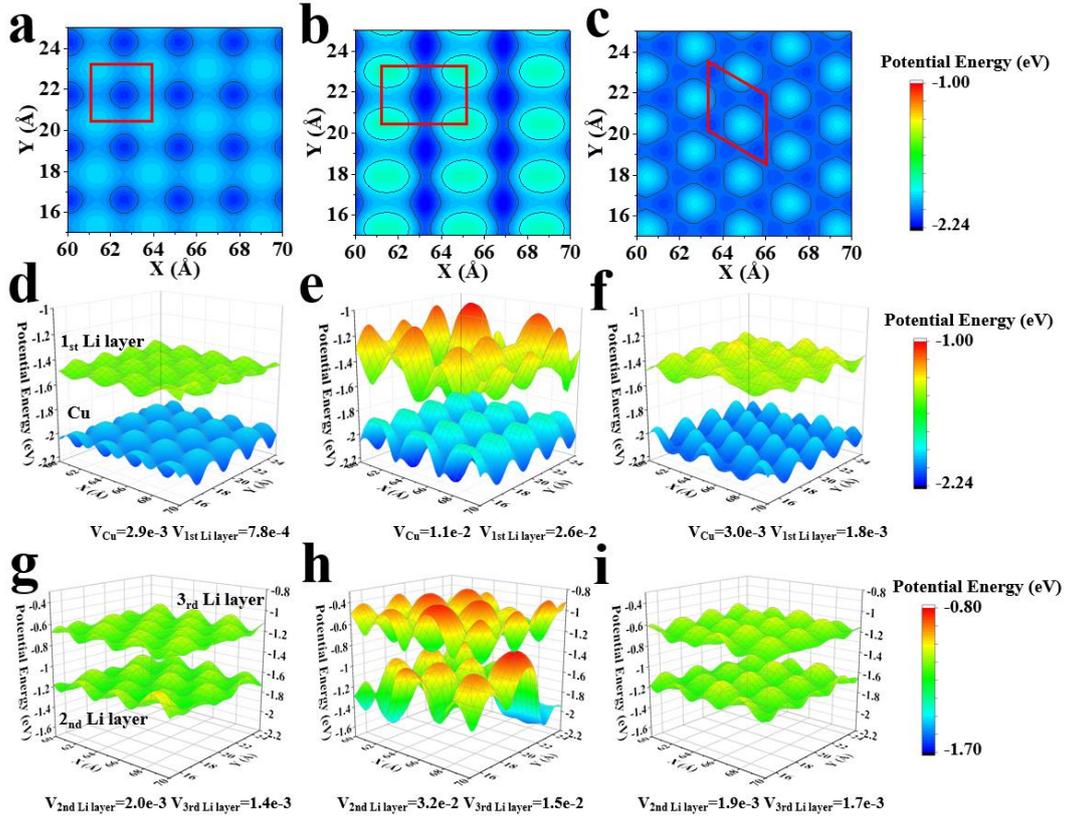

**Fig. 3** (a-c) The 2D potential energy distribution of Li atoms on the Cu (100) (a), Cu (110) (b) and Cu (111) (c) surfaces. (d-f) Potential energy surfaces diagram of Li atoms on the Cu (100) (d), Cu (110) (e) and Cu (111) (f) surfaces (bottom), potential energy surfaces diagram of Li atoms on the first Li layer (top). (g-i) Potential energy surfaces diagram of Li atoms on the second Li layer (bottom) and the third Li layer (top) on the Cu (100) (g), Cu (110) (h) and Cu (111) (i) surfaces.

However, after the Cu (110) surface is covered by one layer of Li atoms, the potential energy variance increases and the potential energy surface becomes more uneven, as shown in the top panel of Fig. 3e, which means that the movement of Li atoms is more restricted and difficult for Li atoms to form the Li (110) structure, spontaneously. On the other hand, the Cu (110) surface has a high lattice match and high similarity with the Li (100) surface, as shown in Fig. 2g,h. Thus, the second layer Li atoms on the Cu (110) surface are approximately arranged with the structural features of the Li (100) surface. After the Cu surfaces are covered by more layers of Li atoms, the potential energy surfaces of the Li atoms on the Cu (100) and Cu (111) surfaces remain flat, and the potential energy surface of the Li atoms on the Cu (110) surface remains large variance and uneven (Fig. 3g-i). Thus, the arrangements of the second layer Li atoms and the subsequent layer Li atoms deposited on the Cu (100), Cu (110) and Cu (111) surfaces would eventually follow the structural features of the the Li (110), Li (100), and Li (110) surfaces, respectively.

## 2.3 The performance comparison of Cu surfaces with different Miller indices

The simulations of Li homogeneous deposition on the polycrystalline Cu surfaces were performed, as shown in Fig. 4a-c. The structures were visualized by OVITO program using the grain segmentation algorithm.[28] The results show that the structure of Li deposition on the Cu (100+111) surface has almost no grain boundaries or defects (Fig. 4b). However, the structures of Li deposition on the Cu (100+110) and Cu (110+111) surfaces have some grain boundaries or defects (Fig. 4a and c). For comparison, we analyzed the structures of Li deposition on the single crystal Cu surfaces, which realize the growth of single crystal Li without grain boundaries or defects (Fig. S8, Supporting Information). These results show that the Cu (100+111) surface can realize a similar effect to the single crystal Cu surfaces. Thus, if the crystal plane of (110) in commercial Cu foil can be modified to (100) or (111), it may promote the dendrite-free Li deposition.

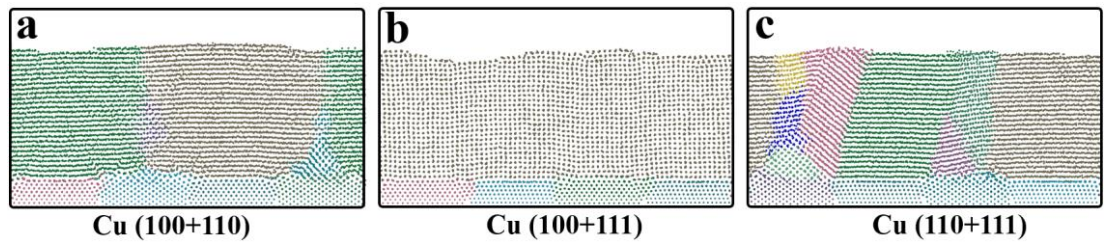

**Fig. 4** Grain analysis of Li homogeneous deposition. (a-c) Grain analysis of Li deposition on the polycrystalline Cu (100+110) (a), Cu (100+111) (b), and Cu (110+111) (c) surfaces.

To further verify the advantages of the Cu (100) and Cu (111) surfaces, we introduced the inhomogeneous deposition to simulate the tip effect on the Cu (100), Cu (110) and Cu (111) surfaces (see Methods section for simulation conditions). Fig. 5 shows the snapshots of the Li atoms spreading on the Cu (100), Cu (110) and Cu (111) surfaces. The Li atoms on the Cu (100) and Cu (111) surfaces spread regularly in the X and Y directions and diffuse quickly to the whole surface in the process of Li initial deposition. Nevertheless, the diffusion of Li atoms on the Cu (110) surface is significantly slower and reflects an intense anisotropy, as shown in Fig. 5e-h. The diffusion velocity of Li atoms on the Cu (110) along the Y-direction is much faster than that along the X-direction. The diffusions of the Li atoms in the X-direction are more difficult than that in the Y-direction due to the anisotropy in the distribution range of the high potential energy sites (Fig. 2f). Moreover, we calculated the energy barriers and mean square displacements of Li atoms on these Cu surfaces, the results are consistent with the above analysis and show that the Cu (111) surface is much better than the Cu (100) surface in the aspect of the mobility of Li atoms (Figure S9-S10, Supporting Information). Due to the fact that the higher fluidity of the deposited Li atoms would greatly facilitate the dendrite-free Li deposition,[21] these results indicate that the Li atoms deposition on the Cu (100) and Cu (111) surfaces have better performance than on the Cu (110) surface in the initial deposition process.

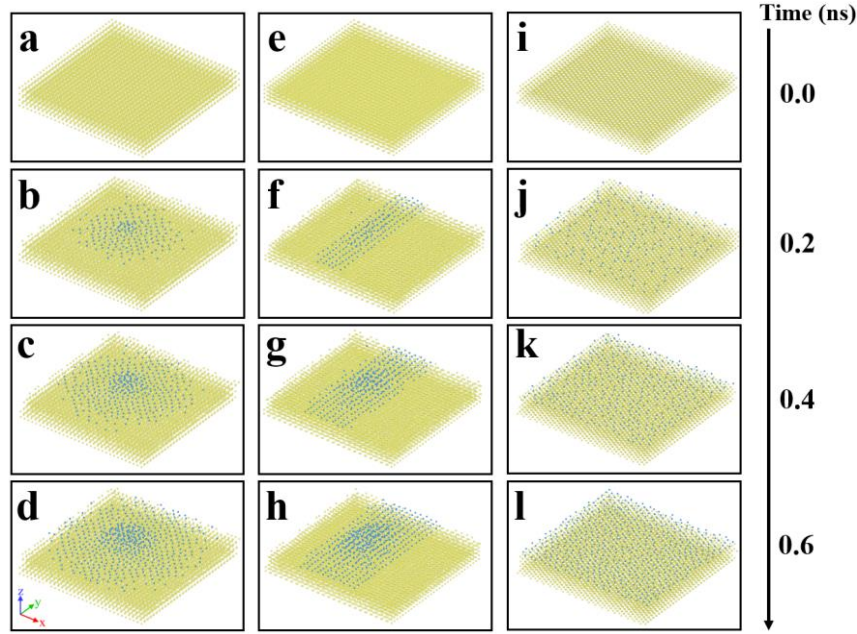

**Fig.5** Snapshots of inhomogeneous deposition at 0-0.6 ns on the Cu (100) (a-d), Cu (110) (e-h) and Cu (111) (i-l) surfaces.

## 3.Disscusion

Our simulations are consistent with many previously reported results. For instance, a study of computational screening of CCs showed that increasing the fraction of the (111) facet on the Cu substrate surface could induce more uniform and dendrites-free Li deposition.[19] Researchers have also found an extraordinary dendrite-free Li deposition on highly uniform facet wrinkled Cu (100) substrates in the experiment.[25] In addition, an experimental investigation of Cu (100) surface induced by electrochemical faceting demonstrated that the Cu (100) surface is lithiophilic and enables Li deposition via the Li (110) oriented Li underpotential deposition layer that guides the growth of Li along the Li (110) direction,[24] which is in good agreement with our simulation results. Combined with these experimental works, we further infer that the Miller index of the Cu surfaces is an important factor in affecting the behavior of the Li deposition on the Cu surfaces. The arrangements of the Li atoms deposited on the Cu surfaces with the different Miller indices have different features and rules. By regulating the exposed surface of Cu foils, such as reducing the fraction of the (110) facet in commercial Cu foils, the deposition behavior of Li on the Cu CC can be effectively improved and promote the dendrite-free Li deposition in the anode-free Li metal batteries finally.

## 4. Conclusion

In summary, we have studied the mechanism of Li deposition on the Cu surfaces with the different Miller indices by MD simulations using LiCu-NNIP with quantum-mechanical computational accuracy. The SSA results revealed that the Li layers on the Cu (100) and Cu (111) surfaces both prefer forming the Li (110) surface structure, but the Li layers on the Cu (110) surface prefer forming the Li (100) surface structure. The potential energy surfaces and lattice features were introduced to illustrate the mechanism of Li deposition on the Cu substrates. In addition, the simulations of Li deposition on the polycrystalline Cu surfaces and Li inhomogeneous deposition with the tip effect proved that the Cu (100) and Cu (111) surfaces have better performance than the Cu (110) surface. Thus, the suggestion of reducing the fraction of the (110) facet in commercial Cu foils was proposed to reduce the possibility of Li dendrite growth. This work provides new insights into the design and construction of Cu CC for improving the performance of anode-free Li metal batteries.

## 5. Method

**5.1. Molecular Dynamics Simulations**

All the MD simulations were performed with the parallel Large-Scale Atomic/Molecular Massively Parallel Simulator(LAMMPS).[29] The force field used for MD simulations was the LiCu-NNIP model with quantum-mechanical computational accuracy, which was specially developed for the simulations of Li-Cu interface system. The NVT ensemble with the Langevin thermostat was used in all deposition simulations.[30] A timestep of 1 fs was used in all the simulations. All the configurations of Cu substrates used in Li deposition were supercells of the primitive FCC cell with a lattice constant of 3.62 Å, which is in agreement with the experimental values.[31] All visualizations of the molecular dynamics trajectory were performed with the OVITO program.[28]

Homogeneous Deposition Simulation Conditions: The lattice constants of the supercells used in the homogeneous deposition were about 23 nm×4 nm×20 nm for single crystal Cu substrates and 18 nm×2 nm×20 nm for polycrystalline Cu substrates. About 1 nm thickness of Cu atoms were the substrate and the half of the bottom layers of atoms were fixed for a bulk environment. Li atoms were generated at the top of the supercell, and the probability of atom generation at X and Y coordinates obeyed the distribution of X~U (0, 23) and Y~U (0, 4) for single crystal Cu substrates (X~U (0, 18) and Y~U (0, 2) for polycrystalline Cu substrates), where U stands for uniform distribution. The time interval of atom generation (generation rate) was 1 Li ps$^{-1}$, and the falling speed of Li atoms was 200 m s$^{-1}$. These simulations were performed at 300 K by NVT ensemble with a total deposition time of 12 ns.

Inhomogeneous Deposition Simulation Conditions: The lattice constants of the supercells used in the inhomogeneous deposition were about 8 nm×8 nm×20 nm. Li atoms were generated at the top of the supercell. The generation region was limited to a small region of 3.8~4.2 nm in both X and Y directions. The generation and falling rates) were the same as the homogeneous deposition conditions. The simulation with results depicted in Fig. 5 was performed at 300 K by an NVT ensemble with a total deposition time of 1.5 ns.

### 5.2. Surface Similarity Analysis

An SSA method developed in this work was used to quantitatively analyze the surface structure of Li layers on the Cu substrates. The SSA method, modifying the similarity analysis method for evaluating the similarity of crystalline, disordered and molecular compounds, is based on Smooth Overlap of Atomic Positions (SOAP) and

REMatch kernel.[32-36] Herein, the SSA method was performed for comparing the structure Li layers on the Cu substrates with the surface structures of the standard Li crystal, which are the Li (100), Li (110), Li (111), and Li (211) surfaces.

# Supporting information for

## The mechanism of Li deposition on the Cu substrates in the anode-free Li metal batteries

*Genming Lai[1], Junyu Jiao[1,*], Chi Fang[1], Liyuan Sheng[2], Yao Jiang[3], Chuying Ouyang[3], and Jiaxin Zheng[1,3,*]*

[1]School of Advanced Materials, Peking University, Shenzhen Graduate School, Shenzhen 518055, People's Republic of China

[2]PKU-HKUST ShenZhen-HongKong Institution, Shenzhen 518055, People's Republic of China

[3]Fujian science & technology innovation laboratory for energy devices of China (21C-LAB), Ningde 352100, People's Republic of China

**Corresponding Author**

* Junyu Jiao, E-mail: jyjiao@pku.edu.cn

* Jiaxin Zheng, E-mail: zhengjx@pkusz.edu.cn

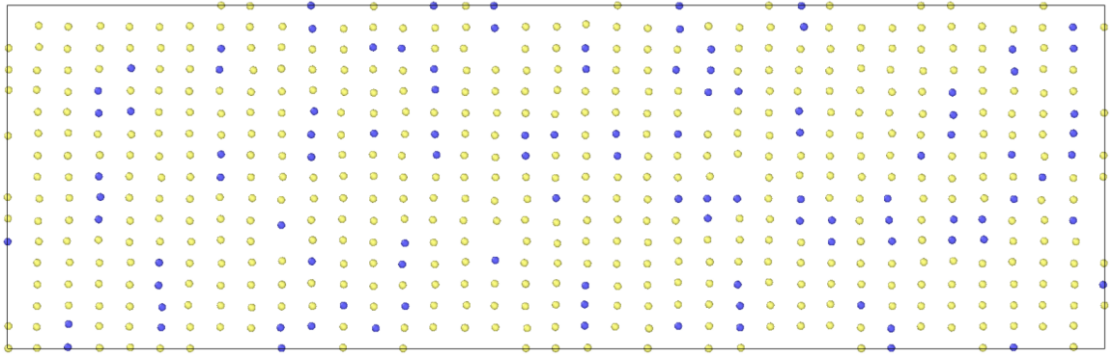

**Fig. S1** Atomic simulation diagram of the Cu (110) surface after deposition.

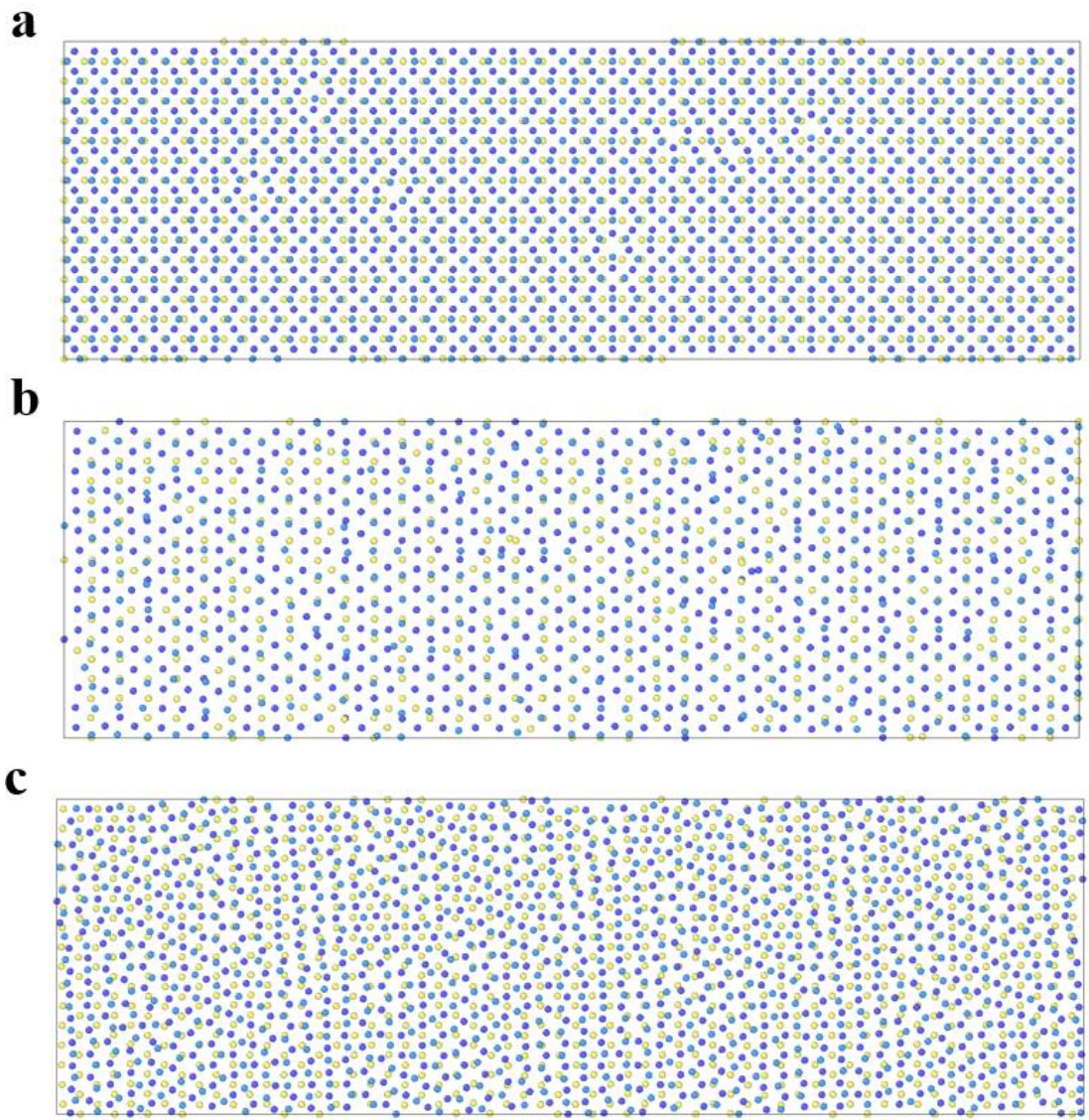

**Fig. S2** The arrangement of the first two layer Li atoms on the Cu (100) (a), Cu (110) (b), and Cu (111) (c) surfaces.

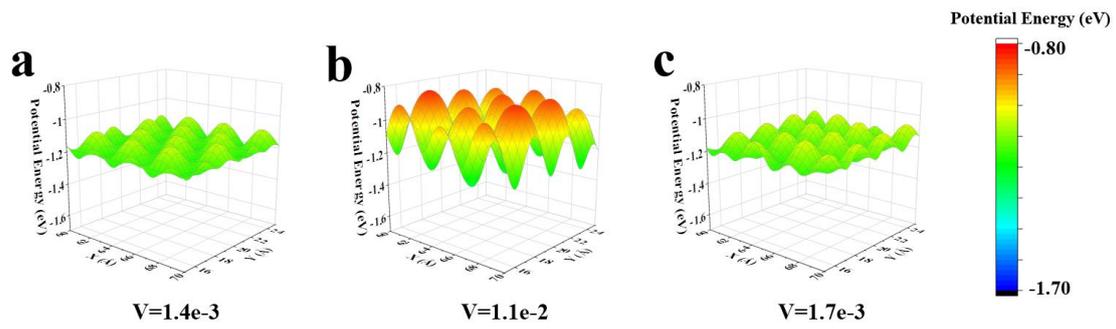

**Fig. S3** The potential energy surface diagram of Li atoms on the Cu (100) (a), Cu (110) (b) and Cu (111) (c) surfaces after covered with four layer of Li atoms.

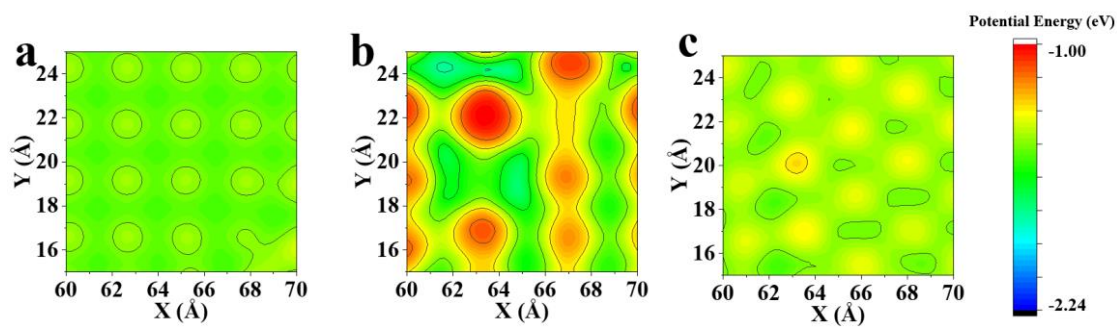

**Fig. S4** The 2D potential energy distribution of Li atoms on the Cu (100) (a), Cu (110) (b) and Cu (111) (c) surfaces after covered with one layer of Li atoms.

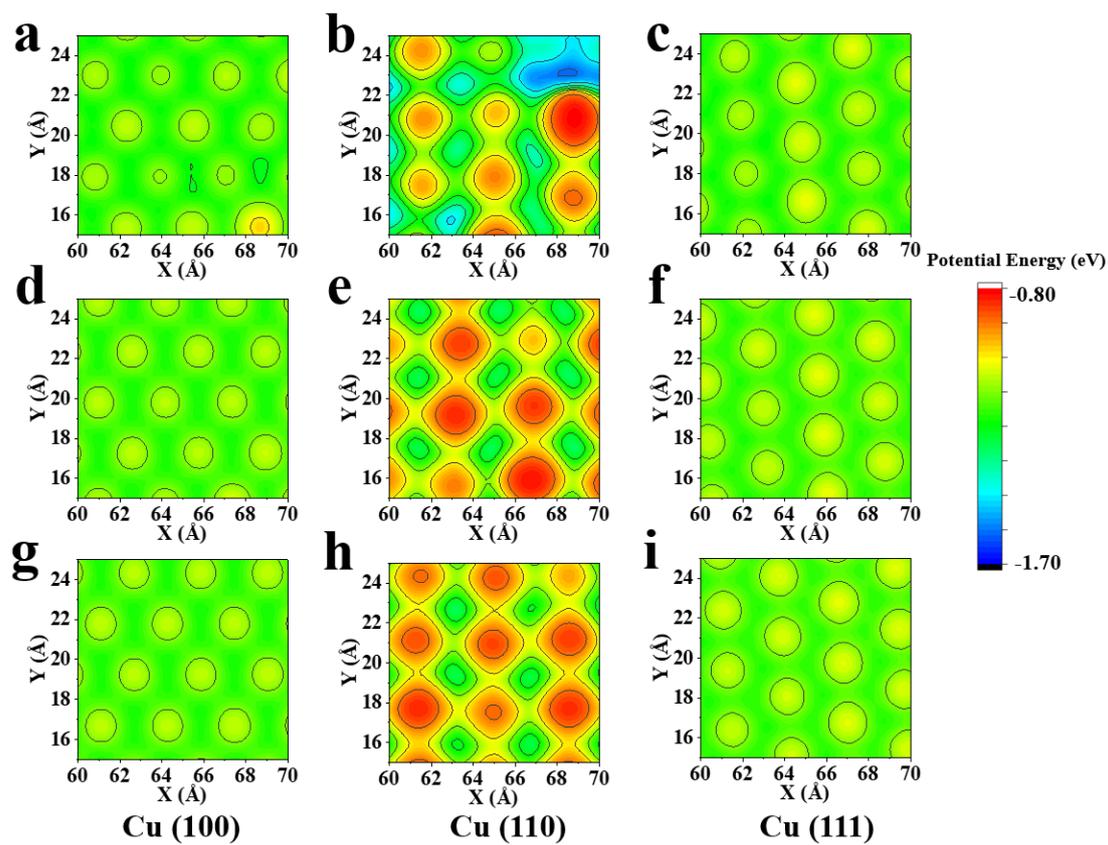

**Fig. S5** The 2D potential energy distribution of Li atoms on the Cu (100) (a,d,g), Cu (110) (b,e,h) and Cu (111) (c,f,i) surfaces after covered with two layer of Li atoms (a-c), three layer of Li atoms (d-f), and four layer of Li atoms (g-i).

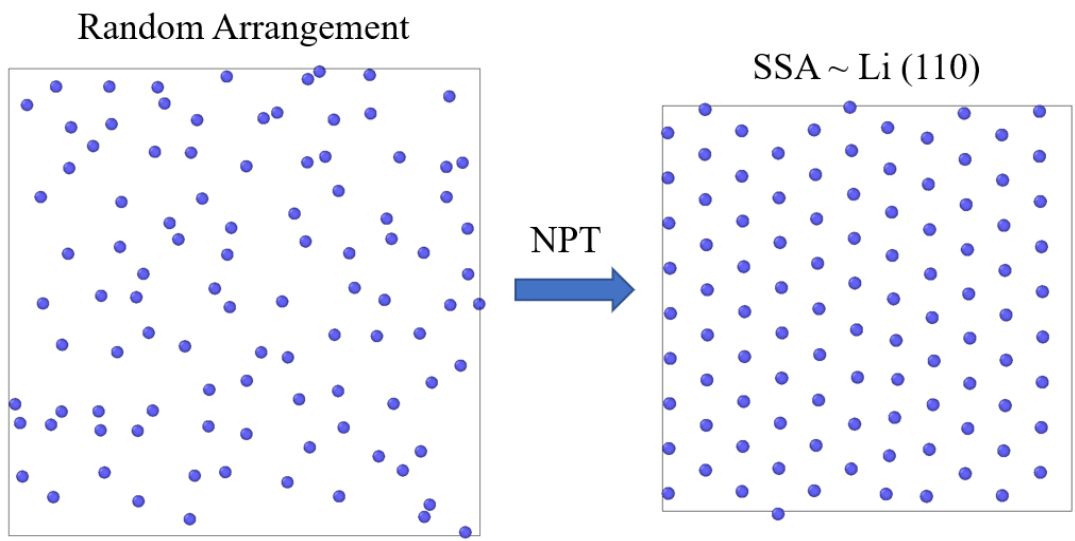

**Fig. S6** The 2D MD simulation of Li atoms. The figure shows that the Li atoms with random initial structure are spontaneously arranged into the structure of Li (110) after relaxation.

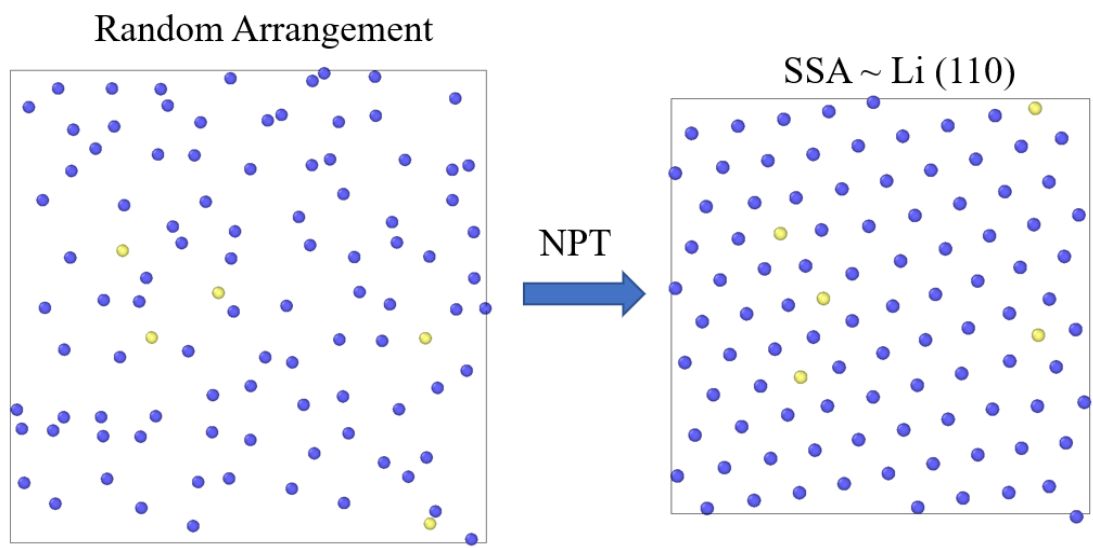

**Fig. S7** The 2D MD simulation of Li atoms and Cu atoms. The figure shows that the initial structure which is randomly arranged with Li atoms and 5% doped Cu atoms, is spontaneously arranged into the structure of Li (110) after relaxation.

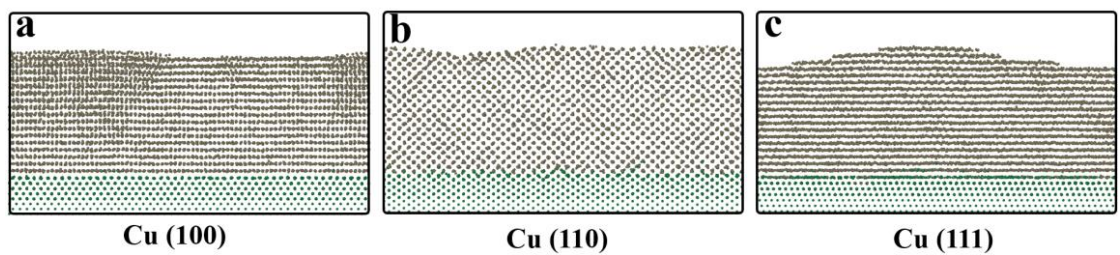

**Fig. S8** Grain analysis of Li deposition on the Cu (100) (a), Cu (110) (b), and Cu (111) (c) surfaces.

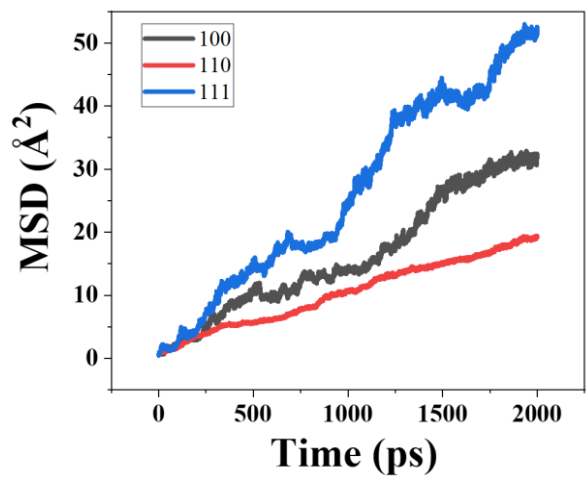

**Fig. S9** Mean square displacement curve of the surface Li atoms on the Li covered Cu (100), Cu (110), Cu (111) surfaces after Li homogeneous deposition in Fig.1.

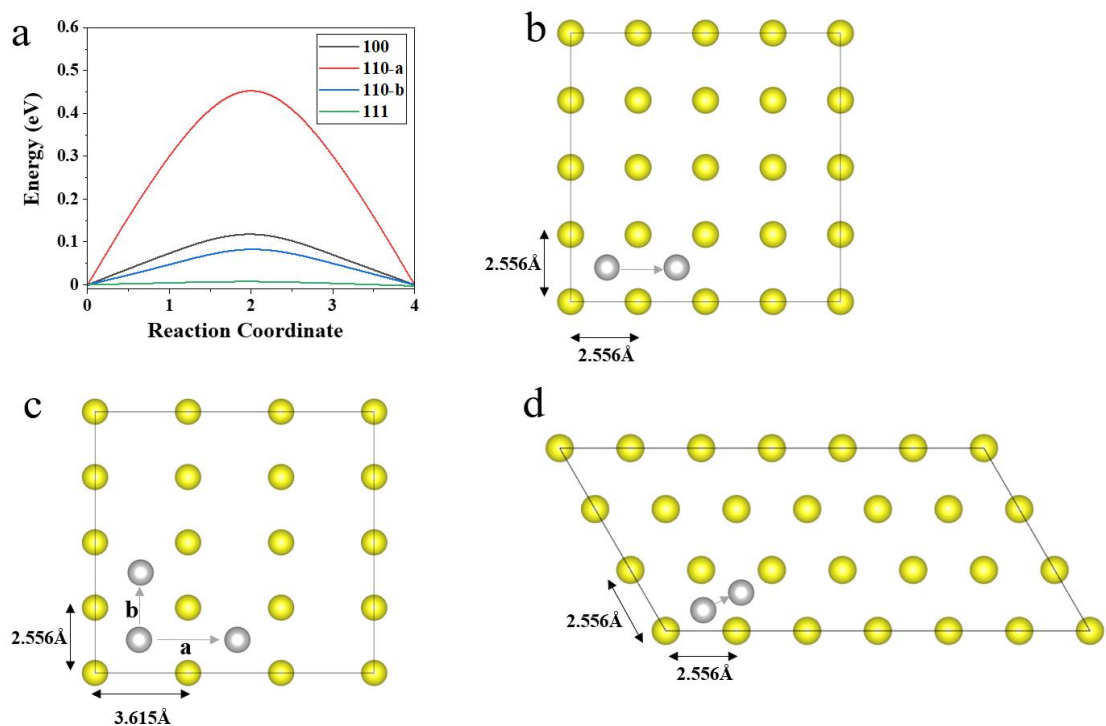

**Fig. S10** (a) Calculated energy barriers of Li atoms on different Cu surface facets. (b-d) Diffusion path of Li atoms on the Cu (100) (b), Cu (110) (c) and Cu (111) (d) surface. These configurations were visualized using the VESTA software.[1] The energy barriers of Li atoms were calculated by using the Vienna ab initio simulation package (VASP) [2,3] together with the climbing image nudged elastic band (CI-NEB) method.[4,5] The generalized gradient approximation (GGA) with a parametrized exchange-correlation function according to Perdew Burke and Ernzerhof (PBE) was used during the calculations.[6] The valence electron wave functions were expanded in the plane wave basis sets, and the projector augmented wave (PAW) method was used to describe the core-electron interactions.[7] The plane-wave cutoff energy was set to be 600 eV.